\documentclass{aastex631}

\shorttitle{ATA--ASP}
\shortauthors{Sheikh and Hort et al.}

\usepackage[nolist]{acronym}
\usepackage[table]{xcolor}

\definecolor{sof}{rgb}{0.5, 0.0, 0.5}

\begin{document}

\title{A Large-Scale Technosignature Survey of the Anti-Solar Point with the Allen Telescope Array: \\ Survey Design and Pilot Results}

\author[0000-0001-7057-4999]{Sofia Z. Sheikh}
\affiliation{SETI Institute, 339 Bernardo Ave, Suite 200, Mountain View, CA 94043}
\affiliation{Berkeley SETI Research Center, 339 Campbell, Berkeley, CA 94720}

\author{Eleanor Hort}
\affiliation{Department of Physics and Astronomy, University of New Mexico, 210 Yale Blvd NE, Albuquerque, NM 87106}

\correspondingauthor{Sheikh and Hort (Co-First Authors)}

\author[0009-0009-1019-3890]{B{\'a}rbara Cabrales}
\affiliation{Jodrell Bank Centre for Astrophysics (JBCA), Department of Physics \& Astronomy, Alan Turing Building, The University of Manchester, M13 9PL, UK}

\author[0009-0003-9610-5068]{Isabel Gerrard}
\affiliation{SETI Institute, 339 Bernardo Ave, Suite 200, Mountain View, CA 94043}

\author[0000-0002-0161-7243]{Wael Farah}
\affiliation{SETI Institute, 339 Bernardo Ave, Suite 200, Mountain View, CA 94043}

\author[0000-0001-9686-5890]{Nick Tusay}
\affiliation{Department of Astronomy, University of Washington, Seattle, WA, 98195, USA}

\author[0000-0003-3197-2294]{David R. DeBoer}
\affiliation{Sub-Department of Astrophysics, University of Oxford, Banbury Road, OX1 3RH}
\affiliation{Radio Astronomy Laboratory, University of California, Berkeley, CA, 94720 USA}

\author[0009-0005-1104-3502]{Saul Balcarcel-Salazar}
\affiliation{Department of Physics, Stanford University, 382 Via Pueblo Mall, Stanford, California 94305, USA}

\author[0000-0002-7735-5796]{Joe S. Bright}
\affiliation{SETI Institute, 339 Bernardo Ave, Suite 200, Mountain View, CA 94043}

\author[0000-0001-5576-2254]{Luigi F. Cruz}
\affiliation{SETI Institute, 339 Bernardo Ave, Suite 200, Mountain View, CA 94043}

\author[0009-0009-6231-9280]{Ben Jacobson-Bell}
\affiliation{Berkeley SETI Research Center, 339 Campbell, Berkeley, CA 94720}
\affiliation{Department of Astronomy, University of California, Berkeley, 501 Campbell \#3411, Berkeley, CA 94720}

\author{Mark Masters}
\affiliation{SETI Institute, 339 Bernardo Ave, Suite 200, Mountain View, CA 94043}

\author[0000-0002-3430-7671]{Alexander W. Pollak}
\affiliation{SETI Institute, 339 Bernardo Ave, Suite 200, Mountain View, CA 94043}

\author{Seth Shostak}
\affiliation{SETI Institute, 339 Bernardo Ave, Suite 200, Mountain View, CA 94043}

\author{Carol Shumaker}
\affiliation{SETI Institute, 339 Bernardo Ave, Suite 200, Mountain View, CA 94043}

\author[0000-0003-2828-7720]{Andrew Siemion}
\affiliation{SETI Institute, 339 Bernardo Ave, Suite 200, Mountain View, CA 94043}
\affiliation{Sub-Department of Astrophysics, University of Oxford, Banbury Road, OX1 3RH}
\affiliation{Berkeley SETI Research Center, 339 Campbell, Berkeley, CA 94720}
\affiliation{Jodrell Bank Centre for Astrophysics (JBCA), Department of Physics \& Astronomy, Alan Turing Building, The University of Manchester, M13 9PL, UK}
\affiliation{Department of Physics, Faculty of Science, University of Malta, Msida MSD 2080, Malta}

\begin{abstract}

In the search for technosignatures, it can be useful to narrow down the cosmic haystack by prioritizing time-coordinate ``Schelling Points'' at which a beacon may be more likely to be transmitted. One such Schelling Point is the \ac{ASP}, the place and moment at which Earth would lie directly between a hypothetical transmitter in the Earth Transit Zone (i.e., ecliptic) and the Sun. In this work, we describe a campaign to survey the \ac{ASP} for radio technosignatures using the Allen Telescope Array, report the results of three initial nights of data from a pilot study in Summer 2023, and discuss the future of the campaign. In the pilot study, we find no sky-localized narrowband 1--2.4 GHz technosignatures within an RA range of approximately 20.266--21.312~hrs during the \ac{ASP}, setting an upper limit flux density threshold of 115--936~Jy.

\end{abstract}

\acresetall

\section{Introduction} 
\label{sec:intro}

\subsection{The Anti-Solar Point}
\label{ssec:asp}

``Technosignatures'', astronomically-observable phenomena generated by technologically-capable life, provide a complementary astrobiological search strategy to ``biosignatures'' in the search for life in the universe \citep{wright2020planck}. The microwave frequencies have been historically favoured as a place to search for technosignatures, with a strong scientific basis well-described in the frame of the Nine Axes of Merit for technosignature searches \citep{sheikh2020nine}. Radio waves traverse the galaxy at the speed of light, are more energy-efficient in bit-rate (especially if they have a narrow bandwidth) than higher-energy photons, and they are not attenuated by galactic gas and dust. In addition, the astronomical radio sky is generally quite quiet, with only a few, morphologically distinct classes of natural confounders that look very different than the signals produced by human technology; human radio transmitters tend to emit signals that are orders-of-magnitude spectrally narrower than masers, the narrowest natural phenomena \citep{tarter2001search}. Finally, using radio technology to communicate is something that humans have done for approximately a century, meaning that there is no extrapolation needed to unknown technologies. 

In technosignature searches, we often make an assumption that we are looking for an intentional communication that is being broadcast by an \ac{ETA} to interstellar space\footnote{Some recent papers avoid this assumption, such as the planet-planet occultation search in \citet{tusay2024ppo}.}. For radio searches specifically, this assumption is broadly driven by the fact that even the strongest omnidirectional ``leakage'' radiation from the Earth \citep[i.e., 4G LTE transmissions, ][]{saide2023simulation} is thousands of times weaker than our strongest directional transmission capabilities on Earth \citep{ekers2002seti}. 

Even after the simplified assumption of an intentional beacon, there is still an immense 8-dimensional parameter space to search \citep{wright2018much}. Thus, we additionally assume that energy conservation and efficiency are important to the \ac{ETA} sending the signal. There are two general ways to conserve energy in interstellar radio signaling, assuming a fixed morphology: 1) send signals only intermittently, instead of persistently 2) beam the power in a specific direction (instead of building an omnidirectional transmitter) \citep{corbet2003synchronized}. The issue with these low duty-cycle, beamed signals is that they are unlikely to be picked up from sky surveys by chance; while upcoming all-sky, all-the-time programs such as PanoSETI \citep{maire2022panoramic} may have a chance of serendipitously observing them, no such programs exist today, and none exist at radio wavelengths.

To mitigate the detectability issue with beamed, low duty-cycle transmissions, we can rely on the idea of mutually-determinable ``Schelling Points'' \citep{schelling1980strategy} from the game theory literature, which has made interdisciplinary inroads to the technosignature literature \citep{wright2020planck}. To apply this theory to an \ac{ETA}'s beamed transmission to Earth, we need to consider which astronomical parameters can be used to signal a mutually-determinable \textit{time} and \textit{coordinate}. Some authors have proposed Schelling points in time synchronized to supernovae \citep{makovetskii1980mutual, davenport2022searching, cabrales2024searching}, nearby variable stars \citep{mclaughlin1977timing}, or gamma ray bursts \citep{corbet1999use}.

Here, instead, we wish to consider a characteristic of our own solar system visible at astronomical distances, with the assumption that Earth is the intended recipient of the signal. If the transmitter happens to be located in the \ac{ETZ}, defined by a narrow range of angles above and below the ecliptic, then the \ac{ETA} would be able to see the transit of the Earth across the face of the Sun \citep{castellano2004visibility}. Transits are an excellent feature to use for a Schelling coordinate because they are easily detectable, they provide extra information about our planetary system (e.g., existence of a planet in the habitable zone, opportunities for planetary atmospheric characterization), and they give clear opportunities for temporal synchronization \citep{sheikh2020breakthrough}. 

With the Earth's transit as our chosen feature, one potential Schelling time is defined when the Sun is directly opposite the transmitter in the sky, i.e., when the transmitter sees Earth at the midpoint of its transit: from here on, we shall refer to the apparent position (from Earth) of this alignment as the \ac{ASP}. This idea was first articulated, for Jupiter's transits specifically, by \citet{singer1982look}, but Earth may be preferable due to its transit-derivable presence in the habitable zone \citep{corbet2003synchronized}. The \ac{ASP} Schelling coordinate/time is generated by a simple geometry with no obvious degeneracies, and it is related to the target habitable planet (Earth) directly, with no need to rely on external (as a nearby periodic astronomical system) or unpredictable (as a high-energy explosion) events \citep{corbet2003synchronized}. 

However, there are some caveats to the \ac{ASP} method. Firstly, either the transmitter or the receiver must account for the light travel time that will cause Earth's mid-transit to appear at different times to an observer on Earth and a distant transmitter. As the transmitter, the \ac{ETA} would have to know the exact distance and proper motion of the sun, including the orbit and orbital phase of the Earth \citep{corbet2003synchronized}. \citet{shostak2004scheme} argue that the number of targets in the transmitter's survey is very likely much smaller than the number of potential near-ecliptic stars as seen from Earth, so the onus should be on the transmitter to correct. Given that the \ac{ETA} that is transmitting will probabilistically have arisen much earlier than humans did \citep[i.e., ][]{kipping2020contact}, they may have better technology for planet detection than we do on Earth\footnote{Of course, this improved technology could also reduce their reliance on the transit method entirely}. The critical factor, the distance to the system, could be determined to high precision with a more sophisticated version of a parallax mission analogous to Gaia \citep{prusti2016gaia}. Note that this survey is also sensitive to probes within the solar system \citep{freitas1980interstellar}, from which it is easy to calculate the required time delay.

Despite this strong motivation for an \ac{ASP} search, no radio technosignature campaign has surveyed with the necessary coordinates and timing. As of 2020, the Breakthrough Listen Initiative \citep{worden2017breakthrough}, did not have a single pointing within $0.5^{\circ}$ of the \ac{ASP} within their Open Data Archive\footnote{\url{https://breakthroughinitiatives.org/opendatasearch}; this work was performed by co-author Balcarcel as a final project for the \\ online Introduction to Astronomy Research course, advised by co-author Sheikh \\ \url{https://github.com/howardisaacson/Intro-to-Astro-2020}}. Large radio technosignature surveys such as META \citep{colomb1995results} and BETA \citep{leigh2000strategies} likely spent only 0.0004\% of their survey time on the \ac{ASP}, entirely serendipitously; the optical search performed by \citet{horowitz2001targeted}, at best, would have resulted in 0.05\% \citep{corbet2003synchronized}. A dedicated survey will be necessary to place limits on radio technosignatures leveraging the \ac{ASP} as a Schelling Point.

\subsection{The Allen Telescope Array}
\label{ssec:ata}

The \ac{ATA} is a 42-dish radio interferometer located at \ac{HCRO} in Hat Creek, California. It is operated by the SETI Institute, and was purpose-built for radio technosignature surveys \citep{welch2009allen}, making it the optimal instrument for a long-term \ac{ASP} survey. 

The \ac{ATA}, as a ground-based facility, can only observe the \ac{ASP} when it is above the horizon at Hat Creek, i.e., during the nighttime, which would lead to a total survey coverage of approximately half of the ecliptic over one year (more densely populated during the Northern Hemisphere winter months, and more sparsely populated during the Northern Hemisphere summer months). Due to the $\sim$365.25 day orbit of the Earth and the presence of a leap day every four years, the available swath of the ecliptic as viewed from Hat Creek precesses, such that over four years the ASP will be observable over the entire ecliptic. Covering this entire region would result in a four order-of-magnitude sky coverage improvement for this technique.

Each 6.1~m element of the \ac{ATA} is being retrofitted with cryocooled, log-pyramidal ``Antonio'' feeds, which are sensitive from 1--10~GHz. Currently, 28 of the 42 antennas have been outfitted with the new feeds and are used as part of the high spectral resolution beamformer backend that is being deployed on the \ac{ASP} survey. Within the 1--10~GHz range, the science user can tune the on-site local-oscillators to any four 672~MHz spectral windows for observation; note that only two were operable until Summer 2025, and the initial survey design maintains the two-tuning setup.

In Section~\ref{sec:survey_design}, we will discuss the observing strategy, analysis strategy, and upper limits that will be set by the ATA-ASP survey. In Section~\ref{sec:pilot_obs}, we discuss the results from the three nights of pilot observations with the ATA in Summer 2023, setting some initial limits on a few discrete \ac{ASP} pointings. Finally, we will discuss the future of the ATA-ASP survey and conclude in Section~\ref{sec:conclusions}.

\section{Survey Design}
\label{sec:survey_design}

\subsection{Observing Strategy}
\label{ssec:observing_strats}

If the \ac{ATA} were dedicated to ATA--ASP for 100\% of its observing time, the available half of the ecliptic would be covered exactly one year after it took its first night of data. Two years later, the campaign could be reperformed to cover the other half of the ecliptic. However, other projects will also be in the nightly observing queue, especially priority sources for other \ac{ATA} science areas such as fast transients \citep[e.g., \acp{FRB}][]{sheikh2024characterization} and slow transients \citep[e.g., supernovae][]{bright2023precise}. Given that, we expect ATA--ASP to be on-sky approximately 2--3 days per week, and for the survey to therefore take about four years to obtain full coverage of the ecliptic.

For this survey, we are utilizing the \ac{ATA}'s high spectral resolution beamformer, which runs on the \texttt{\ac{BLADE}} software\footnote{The software is available on GitHub \texttt{BLADE} codebase: \url{https://github.com/luigifcruz/blade} under an MIT License. \\ The version used for this analysis is archived in Zenodo \citep{blade_cruz_2024}.}. The beamformer produces a \texttt{filterbank} HDF5 ``data cube'' of time, frequency, polarization, and flux data corresponding to particular points within a single dish's field-of-view. Note that we do not perform any coherent Doppler correction (e.g., to the solar system barycenter) before production of the data cube, but instead incoherently correct for drift rates in the \texttt{bliss} stage of the analysis pipeline (Section~\ref{sssec:bliss}). Coherent de-Doppler methods allow for more power to be recovered for drifting signals, as discussed in \citet{margot2021search}, but is computationally expensive; we plan to develop and deploy de-Doppler methods on \ac{ATA} eventually, but they are not used in this work.

Technosignature searches can in theory be conducted in the image plane as well, giving certainty to the localization of emission on the sky, but that strategy is still under development for \ac{ATA} data. To ensure sky-localization in a beamforming mode, we instead form two simultaneous synthesized beams on the sky: one at the location of the \ac{ASP}, and one a suitable number of synthesized beamwidths away, but still within the \ac{FWHM} of the primary beam \citep[as employed by][]{tusay2024ppo}. This is similar to the ``nodding'' or ABABAB strategy used by single-dish technosignature searches such as \citet{choza2023breakthrough}. For an array, signals which a) appear in the target ``on-beam'' and b) are significantly attenuated in the control ``off-beam'' are candidates for signals which are sky-localized: distant enough from Earth that they appear as point sources.

We chose an observation time of 5 minutes for each coordinate pointing on the \ac{ASP}. This duration is short enough that the \ac{ASP}'s precession should not significantly affect its position within the beamformed beam, but long enough to be able to distinguish drifting signals. It also creates output files that are manageable in data volume, despite requiring extremely fine sampling in frequency. The frequency resolution in these data is 1.9~Hz, with a time resolution of 16.8~s. The 1.9~Hz sampling is sufficient to resolve narrowband (Hz-scale) signals; combined with the 16.8~s integration time, physically-meaningful drift rates are easily identifiable in the dynamic spectra.

As described in Section~\ref{ssec:ata}, the \ac{ATA} is capable of recording up to $\sim$2.8~GHz of simultaneous bandwidth, divided into four independently movable 672~MHz chunks. For this survey, to keep data rates manageable with the high observing cadence, we select and record with only two of these tunings. While L-band (1--2~GHz) was favored in early \ac{SETI} experiments \citep[because it contains the ``Water Hole'', ][]{machol1975interstellar}, modern radio technosignature searches have been expanding through the radio spectrum, with increasing amounts of observations taken in the 1--12~GHz range by \ac{BL} \citep[e.g., ][]{sheikh2020breakthrough}. If we assume a uniform prior across frequencies \citep[not relying on ``magic frequencies,'' ][]{zuckerman1985preferred} and note that no frequency has been thoroughly explored as of yet in a parameter space or ``cosmic haystack'' sense, then practical considerations will dictate our choice of frequencies. The two practical considerations that drove our frequency selection were 1) beamsize 2) \ac{RFI}. 

\citet{wells2018transit} estimates the \ac{ETZ} to have a width of 0.53$^{\circ}$. The \ac{ATA} has a large primary beam size, such that the primary beam covers the entire \ac{ASP} for any frequency below 6.5~GHz. However, the synthesized or beamformed beamwidth is smaller than the primary beamwidth by a factor of approximately 50 --- there is no \ac{ATA}--observable frequency for which the synthesized beam will cover the entire \ac{ASP}. At such high frequency resolutions, and, commensurately, data rates, the \ac{ATA} can currently record up to two simultaneous synthesized beams. Thus, to cover as much of the \ac{ASP} as possible, lower frequencies (resulting in larger synthesized beams) are preferred. 

However, the second consideration is \ac{RFI}, which is notoriously bad (especially at L-band) and only worsening \citep[e.g., ][]{schirru2023radio}. To work around the worst regions of \ac{RFI}-contamination at \ac{HCRO} while still keeping the synthesized beam as large as possible, we selected center frequencies of 2692~MHz and 3364~MHz, resulting in a frequency range of 2356--3700~MHz.

To further improve the robustness of the survey to \ac{RFI} from ubiquitous satellite communication constellations, we have implemented telescope boresight avoidance (TBA, \citealt{2024ApJ...971L..49N}).  TBA is implemented by sharing the observing location with the satellite operator (in the first instance the direct-to-cell service enabled by the SpaceX Starlink satellites) and they moderate the transmission by either limiting beamforming or turning off the transmitter as the transmitting satellite is near or traversing the main beam, respectively. The direct-to-cell signal occupies the band of 1990--1995~MHz, which is outside of the band in this work, however, the signal strength is such that without this technique the entire RF band of the ATA is saturated and unusable during the roughly 15 seconds of the traverse. Currently, for a ten-hour observation, TBA is applied to mitigate direct-to-cell in, on average, 50 independent instances.

\begin{figure}
    \centering
    \includegraphics[width=\linewidth]{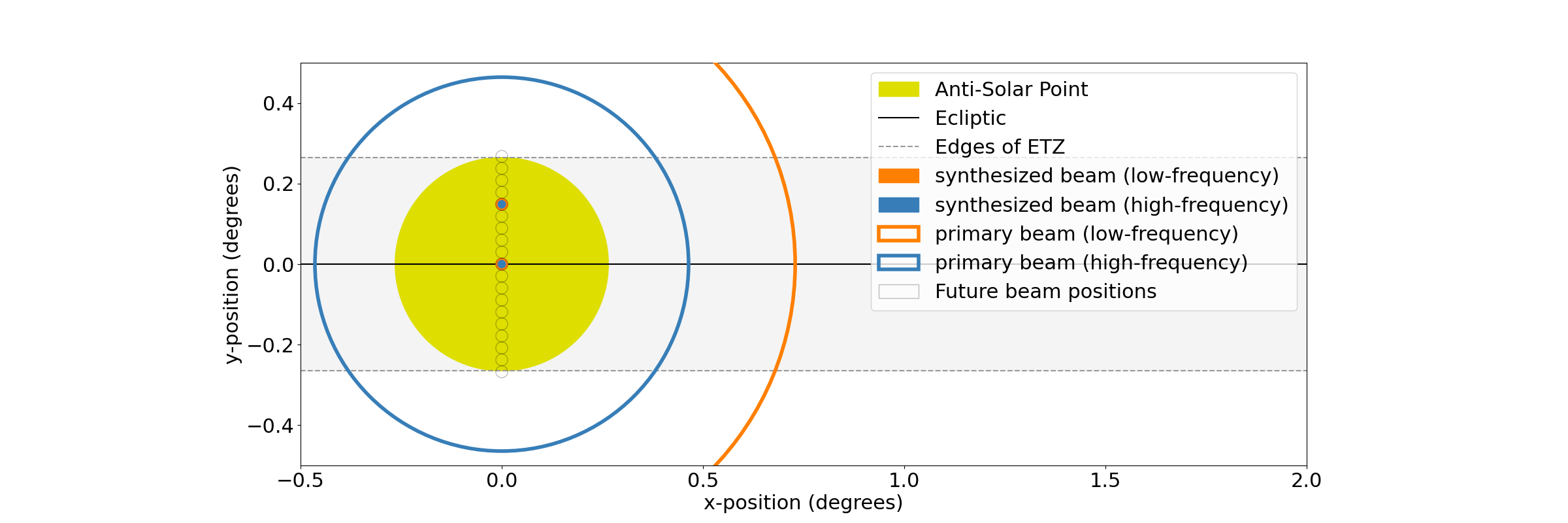}
    \caption{The on-sky geometry of the ATA-ASP campaign. The black horizontal line represents the ecliptic, with the gray shaded region indicating the edges of the \ac{ETZ}. The size of the \ac{ASP} at any one moment is displayed in yellow, with the lowest and highest frequency primary beamsizes overplotted for scale (red and blue circular outlines, respectively). The synthesized beamsizes and positions for the lowest and highest frequencies are shown as shaded red and blue regions respectively within the \ac{ASP}. As the survey continues, we may add more beams in the positions indicated by the gray circular outlines within the \ac{ASP} or shift the survey to the image-plane, where the imaged region would be represented by the primary beam sizes. It should be noted, however, that the initial 2-beam survey is still covering a fully uncharted region of technosignature parameter space.}
    \label{fig:ASP_geometry}
\end{figure}

The geometry of the observing scheme is shown in Figure~\ref{fig:ASP_geometry}. We place one synthesized beam on the center of the \ac{ASP} (aka. the region directly on the ecliptic), which corresponds to observers that see the Earth transit the Sun with the lowest impact parameters. This covers the innermost $\sim$5\% of the \ac{ASP}'s diameter: 0.030$^{\circ}$ at 2356~MHz and 0.019$^{\circ}$ at 3700~MHz. The other synthesized beam is reserved for the ``off-beam,'' and is located five synthesized beamwidths (at the lowest frequency in the band) from the on-beam, perpendicular to the ecliptic. Five beamwidths at the lowest frequency in the survey is equivalent to $\sim8$ beamwidths at the highest frequency in the survey; this is large enough to get isolation between the two synthesized beams at the lowest frequencies, but also small enough that the off-beam will still be comfortably located inside the primary beam and the \ac{ASP} even at the highest frequency in the band. As the \ac{ASP} sweeps across the ecliptic, target stars enter and leave the region (horizontally, in the geometry of Figure~\ref{fig:ASP_geometry}). Thus, even though the monitored regions from the synthesized beams do not cover parts of the \ac{ASP} to the left and right of the central vertical bisector, all targets in those parts will cross the bisector eventually (or have already crossed it). During an observing session, we loop the five minute scans of the \ac{ASP} (with a variable $\lesssim$60~seconds downtime between scans) for the entire time that the \ac{ASP} is above the horizon; this means that we will always have an ecliptic longitude overlap between scans of over 80\%, even at the highest frequencies in the survey. We note that this provides far more than five minutes of coverage for each actual sky location, and that any true signals that we see could show up in multiple contiguous pointings (although this would depend on the duration of the transmission, so we do not rely on it in our sensitivity calculations).

We can also assume that any \ac{ETA} using the \ac{ASP} technique would prioritize their transmission timing to align at mid-transit (i.e., on the bisector). This implies that the total ``coverage'' of the \ac{ASP} in this survey will eventually be 100\% (or 360$^{\circ}$) in ecliptic longitude and 10\% in ecliptic latitude, as 10\% is the ratio of the \textit{diameter} of the \ac{ASP} to the (sum of the) synthesized beams. We aim to improve this coverage in future epochs --- see Section~\ref{sec:conclusions} for a technical review of available options.

\subsection{Analysis Strategy and Event Filtering}
\label{ssec:analysis_strats}

\begin{figure}
    \centering
    \includegraphics[width=\linewidth]{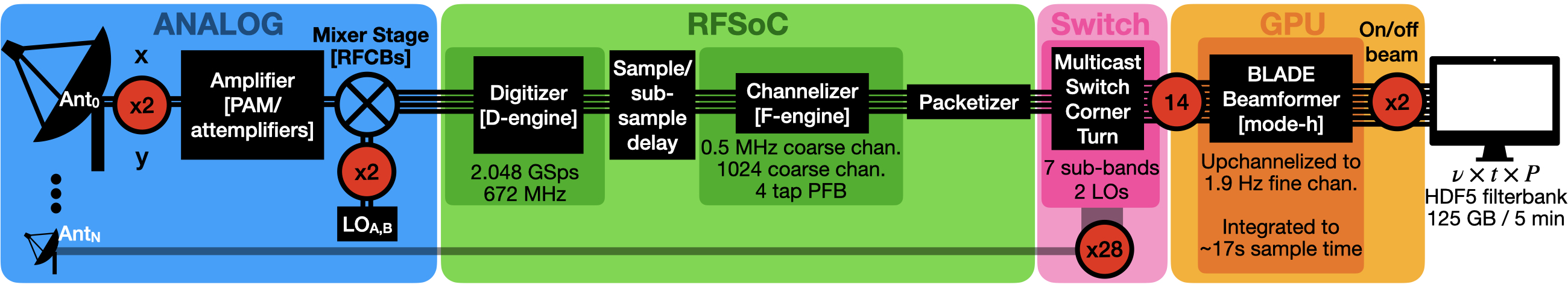}
    \caption{The analog and digital signal path of the \ac{ATA} as used for this work.}
    \label{fig:block}
\end{figure}

 The full block diagram of the analog path and digital signal processing chain for this project is shown in Figure~\ref{fig:block}. For each 5-minute scan of ATA-\ac{ASP} data, the system will output 28 individual \texttt{filterbank} HDF5 files: each of the 2 tunings is divided into 7 frequency chunks, and there are 2 beam files (on and off) in each chunk. One of these files is 4.5~GB in size. With an average of about 80 scans per night, over 365 nights, this project would generate $\sim$3.7~PB of total data. This is approximately the storage capacity of the entire \ac{HCRO} server; it is clearly infeasible to store and process all of this data in a fully offline fashion, and thus a pseudo-realtime processing strategy must be implemented.

The general processing pipeline is as follows: narrowband ``hits'' (Hz-scale signals with drift rate and \ac{SNR} values within the search range) will be found by the \texttt{bliss} code\footnote{The software is available on GitHub \texttt{bliss} codebase: \url{https://github.com/UCBerkeleySETI/bliss} under a 3-Clause BSD License. \\ The version used for this analysis is archived in Zenodo \citep{nathan_west_2025_bliss}.}. These hits will be filtered with an \ac{RFI}-mask and drift rate threshold. Then, the remaining hits will be cross-referenced between the two beams using the \texttt{NBeamAnalysis}\footnote{The software is available on GitHub \texttt{NBeamAnalysis} codebase: \url{https://github.com/isabelgerrard/NBeamAnalysis} under a \\ 3-Clause BSD License, and version 2.0 is archived in Zenodo \citep{ntusay_igerrard_nbeam}.} ``event'' finding code \citep{tusay2024ppo}, which searches for sky-localized hits that are significantly brighter in one beam than another. Any remaining events after the \texttt{NBeamAnalysis} filtering will be inspected visually. If no sky-localized narrowband signals are found in a session, the original \texttt{filterbank} data for that night will be deleted while the reduced data products---CSV files of hits and events, logs, and output plots---will be saved permanently. Conversely, if a signal-of-interest (a potentially sky-localized narrowband signal) is found in a session, both the original data and the output products described above will be saved permanently, for the entire session across all frequencies, to \ac{HCRO} storage. From the pilot observations and other preliminary tests, we find that signals-of-interest are rare enough that this procedure ensures that the data rate remains manageable while keeping full spectral and temporal context around interesting hits.

\subsubsection{\texttt{bliss}}
\label{sssec:bliss}

In general, recent narrowband technosignature searches have used either the \texttt{turboSETI} code \citep{enriquez2019turboseti}, developed by \ac{BL}, or the UCLA SETI pipeline \citep{margot2023search}. On the \ac{ATA}, only \texttt{turboSETI} had previously been implemented. \texttt{turboSETI} estimates drift rates by ``dedrifting'' a signal according to a set of trial drift rates and choosing the one that maximizes the dedrifted signal's amplitude when the dedrifted spectrogram is summed over time. The method is computationally intensive, but is aided by a fast tree algorithm originally developed by \citet{Taylor74} for pulsar dedispersion. A GPU-accelerated version of the turboSETI code, called \texttt{seticore}, is available,\footnote{The software is available on GitHub \texttt{seticore} codebase: \url{https://github.com/lacker/seticore} under an MIT License.} and further details of the method are given in \citet{Enriquez17}.

\texttt{bliss} performs a direct dedrifting routine, i.e., not using the tree algorithm of \citet{Taylor74}. This results in greater computational complexity, but mitigates the dedrifting loss characterized by \citet{gajjar2021breakthrough} see Section~\ref{ssec:upper_limits} for more details. After dedrifting the spectrogram at each trial drift rate, \texttt{bliss} sums over time to obtain a 1D spectrum. All dedrifted spectra are stored sequentially in a drift--frequency plane, analogous to the dispersion measure--time ``butterfly plot'' plane used in some searches for pulsars and fast radio bursts \citep[e.g.,][]{Foster18}. \texttt{bliss} executes its hit search in this plane, which offers several advantages in signal parameter estimation.

It has been recently appreciated that \texttt{turboSETI} has some drawbacks. \citet{choza2023breakthrough} found that the rolloff from standard coarse channelization causes \texttt{turboSETI}'s noise statistics to be skewed, returning a lower estimated mean and higher standard deviation than it should with proper rolloff mitigation. Signals can also be attenuated by tens of percent if they fall near the edge of a fine channel (instead of at its center; see discussion in Section~\ref{ssec:upper_limits}), and can be attenuated by factors of a few to several without additional mitigation for Doppler smearing at drift rates beyond the ``unit drift rate.'' Overall, \citet{choza2023breakthrough} discovered with injection-recovery testing that a user-input \ac{SNR}-threshold of 10 actually corresponds to an \ac{SNR}-threshold of 33 times the true noise in the dynamic spectrum in a \ac{GBT} observation.

\texttt{bliss} accounts for signal attenuation due to both coarse-channel rolloff and Doppler smearing at high drift. The shape of the rolloff is deterministic and, given certain parameters of the signal processing pipeline, its mitigation is as straightforward as modeling the shape and dividing it out of the data in the frequency domain. The \texttt{bliss} pipeline performs this correction as a standard part of its hit search. The correction does not take into account the shape of the bandpass, which can still bias the noise statistics and result in false positives even after the rolloff correction. However, this typically only occurs in rolloff regions and is readily taken care of by our RFI masking procedures. Meanwhile, the direct dedrifting approach taken by \texttt{bliss} permits better characterization of Doppler smearing than the tree algorithm, with the result that much of the signal power lost can be recovered, or ``desmeared.'' A forthcoming study by \citet{jacobson-bell2026bliss} conducts signal injection and recovery tests using \texttt{setigen} \citep{brzyckisetigen} to simulate narrowband signals at various widths, drift rates, and SNRs. The authors find that, while \texttt{turboSETI} consistently underestimates SNR by a factor of $>$3.3, and that this factor is worse for higher-drift signals, \texttt{bliss}'s SNR estimation is $\sim$1 for low-drift signals and still an improvement on \texttt{turboSETI}'s at high drifts. Injection-recovery testing through the entire ATA pipeline, including \texttt{NBeamAnalysis}, is currently underway and will be reported in future work \citep{gerrard2026}.

In addition to the SNR underestimation reported by \citet{choza2023breakthrough}, \texttt{turboSETI} has been criticized for its ``top hit search,'' a routine designed to avoid redundant detections that, in practice, results in the discarding of $\sim$kHz-wide spectral windows surrounding high-SNR signals \citep{Pinchuk19, Margot21}. This is a very small fraction of the $\sim$GHz-wide bandwidths typical of SETI surveys, but it nevertheless biases our constraints. By leveraging the full drift--frequency plane at once, \texttt{bliss} can disambiguate nearby signals more effectively. A user-selectable parameter designates the minimum $L_1$ distance between two peaks in this plane for them to be considered separate signals; that is, \texttt{bliss} requires that signals not only align in frequency but also have similar drift rates to be counted as redundant. A search with this parameter is referred to as a ``connected components'' search. Overlapping signals, then, will typically be missed by \texttt{turboSETI}, but found by \texttt{bliss} so long as they have different drift rates.

Another assumption made by \texttt{turboSETI} is that any signals it finds have bandwidths less than the survey's frequency resolution. When \texttt{turboSETI} detects a signal that spans more than one frequency bin, it does not integrate the signal's power across its frequency profile, giving a per-frequency power value that is assumed to be the total power in our calculation of transmitter limits (see Section \ref{ssec:upper_limits}). (Indeed, \texttt{turboSETI} has no way of estimating a multibin signal's bandwidth at all!) In contrast, \texttt{bliss} takes into account that multibin signals are possible and estimates each signal's bandwidth and total power accordingly. \citet{jacobson-bell2026bliss} conduct their injection-recovery tests using simulated signals up to 10 bins wide and find that \texttt{bliss} consistently recovers signals of these bandwidths. In our data, wider signals are possible, but typically exhibit substructure in their spectral profiles such that \texttt{bliss} is likely to detect them as multiple narrow ($<$10 bins) subcomponents.

For the above reasons, we use \texttt{bliss} instead of \texttt{turboSETI} as our narrowband hit-finding code for this search. The direct dedrifting approach contributes to substantially longer runtimes, but our implementation at the ATA still runs faster than the rate of data acquisition.

We are running the post-processing pipeline with an \ac{SNR}-threshold value of 15, which will stay constant throughout the study duration unless the \ac{RFI} environment gets significantly worse. It should be noted that the coarse channel shape of the \ac{ATA} is more extreme than that of the \ac{GBT}, which would lead to an even greater difference between the \texttt{turboSETI} and ``actual'' \ac{SNR}, compared to the factor of 3.3 for \ac{GBT} described above. Therefore, while a \texttt{bliss} \ac{SNR}-threshold of 15 seems higher than some \texttt{turboSETI} searches in the past \citep[e.g., ][]{sheikh2023green}, it is likely actually \textit{more} sensitive in absolute terms than these comparable searches.

In terms of drift rate, \texttt{bliss} also requires a maximum drift rate to use for narrowband hit searching. The maximum drift rate contribution from Earth's rotational motion (which dominates the orbital component at Earth's orbital distance) is 0.11~nHz\footnote{1 nHz is equivalent to 1 Hz/s of drift at 1 GHz. This is a useful unit, as the drift rate imparted by a Doppler acceleration is \\ frequency-dependent.} \citep{sheikh2019choosing}. For surveys that are performing a search of the sky without restricting to known planetary systems, \citet{li2023developing} recommends a maximum drift rate limit of 0.44~nHz, assuming no correction on the part of the transmitter. If we sum these values, we get a maximum drift rate threshold of 0.55~nHz, or $\sim$2~Hz/s at the highest frequency in our survey. To add some extra buffer on these assumptions, we will double that value to finalize a maximum drift rate for the ATA-ASP survey of 4~Hz/s, with the assumption that the survey could recover bright, non-drift corrected signals from targets that may be on the ecliptic but not related to the ASP. Note however, that in the \ac{ASP} schema we can assume that any signal that we receive is being sent from a system that is deliberately sending signals towards the Earth. Therefore, it is more likely (when compared to a standard open signal search) that the transmitter would correct their own drift, and thus signals-of-interest at lower drift rates within the range may be particularly interesting.

\subsubsection{Analysis Pipeline}
\label{sssec:analysis}

Given the considerations in this section, we run the hit-finding code \texttt{bliss} for ATA-ASP with an \ac{SNR} of 15, a drift rate range of $\pm4$ Hz/s, we use coarse-channel shape correction and desmearing correction, and we use the connected components method with an $L_1$ parameter of 7 drift--frequency pixels. The outputs from this operation, after being looped through each 5~minute scan, frequency chunk, and beam will be saved to .csv files, with each row containing information on a particular hit. All hit .csv files will be saved for the duration of the survey, regardless of whether they have a signal-of-interest in them --- this will allow us to assess \ac{RFI} and hit statistics throughout the survey.

At this stage, we expect an average of 47,000 hits per 5-minute scan per beam, given the results from the ATA-ASP observations thus far. A significant amount of these hits are due to \ac{RFI} in densely populated parts of the spectra, generally delineated by \ac{FCC} allocations. We visually inspected dynamic spectra from all \texttt{bliss} hits from the first 5-minute scan in Fall 2024, and spot checks thereafter, to assess which regions needed \ac{RFI} masking, i.e., removing all hits within the designated frequencies. We thus determined the following list of frequency regions, where any hits in the designated ranges would be removed from consideration:

\begin{itemize}
    \item 2355.75--2359.5 MHz
    \item 2484--2488 MHz
    \item 2497.25--2498.5 MHz
    \item 2590--2609 MHz
    \item 2565.75--2585 MHz
    \item 2705--2800 MHz
    \item 3264--3274 MHz
    \item 3341--3343 MHz
    \item 3492.75--3492.85 MHz
    \item 3600--3700 MHz
\end{itemize}

In total, this covers ranges allocated to wireless communication services and geostationary satellite downlinks, among others, and  and removes 69.46\% of the hits associated to \ac{RFI} while only masking 18.94\% of the band.

After the frequency masking is completed, we input the masked hit .csv files into \texttt{NBeamAnalysis}. \texttt{NBeamAnalysis} performs a series of cross-checks between the on- and off- beams, looking for hits that are sky-localized \citep{tusay2024ppo}. \texttt{NBeamAnalysis} first compares the hit .csvs for a given scan, searching for hit entries at near-identical frequencies and drift rates. If one is found, the beam ratio is checked: for a 28-dish array, the ratio between the power in a coherent beam centered on an emitting target compared to the power in the incoherent beam should be about a factor of $\sqrt{N_{ants}} = \sqrt{28} = 5.29$. We can use this as a ``worst-case scenario'' value for comparing two coherent beams: for any true sky-localized signal, given a pair of beams that are placed at least a few synthesized beamwidths away from each other, we would expect a beam ratio of at least 5.29. In reality, this ratio would likely be much greater, and the exact value will depend on where the off-beam falls within the beam pattern of the \ac{ATA}. Additionally, there will be cases in which an on-beam detection will be above the \ac{SNR}-threshold, but the off-beam contains the signal \textit{below} the \ac{SNR}-threshold, which will not be contained in the hit file (and therefore, not identified as \ac{RFI} in this methodology). To identify these situations, further reduce false positives, and rank the level-of-interest of events at the end of the pipeline, \texttt{NBeamAnalysis} contains another dynamic-spectrum based comparison step. The code grabs a snippet of the dynamic spectrum from both beams, centered on the hit in the on-beam. Then, it calculates the noise in each snippet, and the \ac{SNR} of a path corresponding to the drift rate and frequency location of the on-beam hit to obtain the beam ratio from the data directly. It also calculates the dot-product of the two snippets, which is reported as the ``DOT score'' between the two beams. In theory, a bright sky-localized signal should make the two snippets very different and produce a DOT score close to zero, while two identical snippets (more likely for local \ac{RFI}) would produce a DOT score close to 1. 

For the ATA-ASP survey, we run the \texttt{NBeamAnalysis} pipeline on a per-night basis, instead of a full-survey basis, and we modify the module slightly, as we need to process several hours of data from a single observing night as close to real-time as possible. We implemented multi-threaded disk buffer caching and addressed redundant metadata loading to reduce I/O overhead; integrated JIT (\texttt{Numba}) runtime compilation to mitigate computation overhead; and optimized under-utilized CPU parallelization. These changes result in a 92.3\% reduction in runtime, i.e. even for the longest nights of the year, the modified \texttt{NBeamAnalysis} pipeline analyzes and plots final hits within 8 hours. We ask \texttt{NBeamAnalysis} to plot only hits with beam ratio values of at least 5.29, and at most 1000 plots per-night to be visually inspected, limited by the 1000 lowest DOT scores. All generated plots of top hits and .csv files of all events from \texttt{NBeamAnalysis} will be saved for the entire survey to keep up with the incoming data rate. 

In the case where a potential signal-of-interest is found, i.e., an event that appears sky-localized and thus warrants follow-up, the entire night of data will be saved. Subsequent follow-up in parallel \citep[following procedures such as those described in e.g., ][]{sheikh2021analysis, meadows2022community} will occur --- one likely situation is a signal-of-interest appearing sky-localized in one night of data, and then reappearing in another night at a different position (thus proving it to be non-sidereal) or being attributed to a particular \ac{RFI} source. In these cases, the data from the former signal-of-interest will remain saved on disk for illustrative purposes, but the rest of the data from the night will be deleted.

\subsection{Expected Upper Limits and Haystack Fraction}
\label{ssec:upper_limits}

To evaluate the sensitivity of a \ac{SETI} survey with an interferometer, we must determine the sensitivity of each element in the array and then combine them as appropriate for the beamforming mode. Currently, the \ac{ATA} normalizes the voltage noise levels to the same Gaussian distribution prior to beamforming and assigns uniform weights ($1+0j$) to all dishes. In the case of similar (but uncorrelated) noise samples per antenna, the beamformer \ac{SEFD} for polarization $p$ is:

\begin{equation}
    \rm{SEFD}_{beamformer, p}=\frac{N}{\big(\sum_{i=1}^{N}{SEFD_{p, i}^{-0.5}}\big)^2}
\end{equation}

Where the $x$ and $y$ polarizations are thus calculated separately with the equation above and then combined into a total beamformer \ac{SEFD} via 

\begin{equation}
    \rm{SEFD}_{beamformer, total} = \frac{1}{\sqrt{\frac{1}{SEFD_x^2} + \frac{1}{SEFD_y^2}}}
\end{equation}

\begin{figure}
    \centering
    \includegraphics[width=0.8\linewidth]{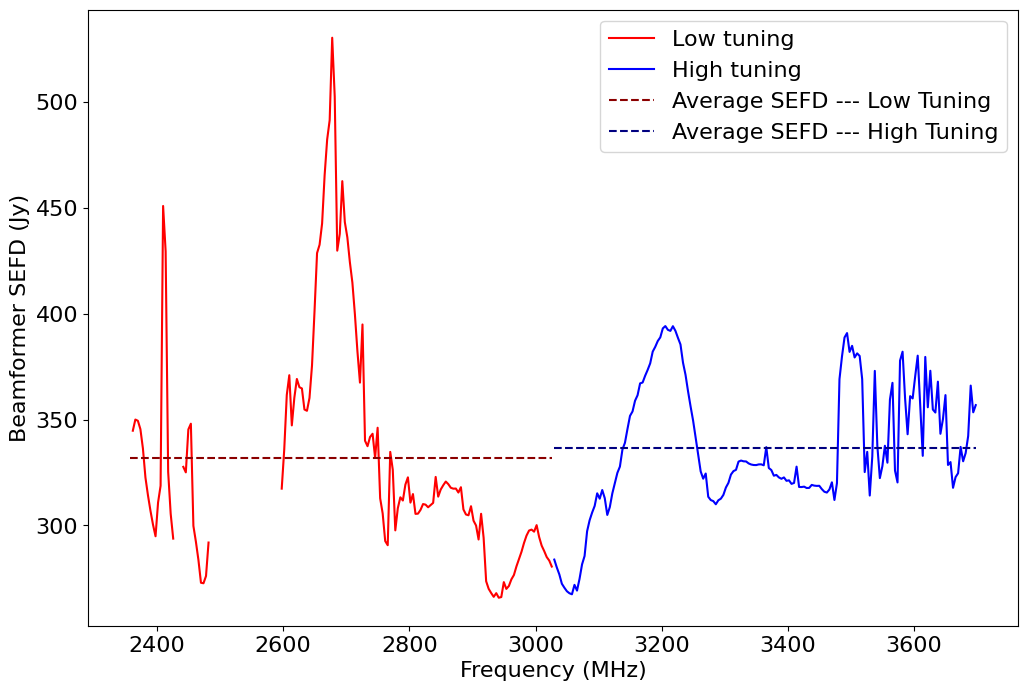}
    \caption{The beamformer \ac{SEFD} in Jy over frequency in MHz for an example night from the ASP survey. The average SEFD is stable across both halves of the bandpass, but large fluctuations are seen in some regions (e.g., $\sim2700$MHz) due to \ac{RFI}. Missing data from $\sim$2500--2600~MHz is also due to \ac{RFI} removal in CASA.}
    \label{fig:sefds}
\end{figure}

We use the calibration from an example \ac{ASP} night (MJD 60893) as a stand-in for the \ac{SEFD} of the full survey: averaging 334~Jy across the entire bandpass. This value does vary across frequency as shown in Figure~\ref{fig:sefds}, but the fluctuations are dominated by \ac{RFI} rather than inherent change in sensitivity from the receiver.

If no signals-of-interest pass verification checks, we will still learn a lot from a non-detection by setting strict new upper limits on radio transmitters using the \ac{ASP} Schelling point, as well as new limits for the \ac{ETZ} in general. The minimum detectable flux density of a narrowband radio technosignature search is modified from \citet{gajjar2021breakthrough} as:

\begin{equation}
    S_{min, narrow} = \eta_{DFT}\eta_{primary}\frac{\mathrm{SNR}_{min}}{\sqrt{\beta}} \frac{\mathrm{SEFD}}{\Delta \nu_t} \sqrt{\frac{\Delta \nu}{n_{pol} \times t_{obs}}}
    \label{eq:minimum_detectable_flux_density}
\end{equation}

where the $\mathrm{SNR}_{threshold}$ for the full ATA-ASP survey is 15, the \ac{SEFD} is $334$~Jy (as used above), the channel width $\Delta \nu$ is 1.9~Hz, the number of polarizations $n_{pol}$ that the search is sensitive to is 2, the observing time per scan is 5 minutes, and we assume a fiducial transmitter signal width of $\Delta\nu_{t,fid}$ = 1~Hz. We can also calculate the minimum detectable flux $F_{min, narrow}$ in e.g., W/m$^2$ by simply multiplying both sides by an transmitter signal width of $\Delta\nu_{t}$. Before plugging values into this equation, we must consider $\eta_{DFT}$, $\eta_{primary}$, and $\beta$ in more detail.

The fine channelization is performed via a \ac{DFT}, which leads to a sinc function channel response as opposed to an ideal boxcar. Due to this, any signal that is significantly larger than the channel width within one time integration (either intrinsically, or due to frequency drift) will recover all of the signal and have a \ac{DFT} efficiency value $\eta_{DFT}=1$. Conversely, an infinitely narrowband, non-drifting signal will have an average $\eta_{DFT}$ of 0.77 \citep[the integral of the sinc function over the channel width, as used in][]{korpela2025seti} and a median $\eta_{DFT}$ of 0.81. In this work, we are looking for narrowband drifting signals with high enough drift rates that we will take $\eta_{DFT}=1$ for our most likely (and most favorable) case, and $\eta_{DFT}=0.8$ for the least favorable case. Note that there is also some maximum detectable flux above which the fine channel will saturate, but we do not address that in this work.

$\eta_{primary}$ accounts for the response of the primary beam at the location of the synthesized beam where the signal is detected. The on-beam is located at the phase center such that $\eta_{primary}=1$. But for the off-beam, assuming a Gaussian beamshape, any signals in the off-beam at the lowest frequencies will have 1\% attenuation, and those at the highest frequencies will have 2.5\% attenuation; in the least favorable case, we thus use $\eta_{primary}=0.975$.

$\beta$ is a parameter representing the ``dechirping efficiency'', which accounts for drift rate smearing past the unit drift rate. For \texttt{turboSETI} (the example in the \citet{gajjar2021breakthrough} paper) $S_{min, narrow}$ scales linearly with $\beta$, but \texttt{bliss} regains some extra signal by including neighboring bins containing smeared signal into the power estimate, reducing the factor to scaling with the square root of $\beta$ (to the first order). The functional form of the dedrifting efficiency is $\beta_{bliss} = [\mathrm{round}\big(\frac{\Delta t}{\Delta\nu} \times |\dot{\nu}|\big)]^{-1}$. Here, $\dot{\nu}$ is the drift rate being searched, which is rounded to the nearest integer multiple of the unit drift rate $\Delta\nu/\Delta t$. 

We make the assumption that \ac{ASP} beacons are likely drift-compensated (see Section~\ref{ssec:analysis_strats}) on the transmitter's side, but that drift contribution from the Earth's rotation and orbital motion will still be present. As calculated in \citet{sheikh2019choosing}, these contributions are at most (with maximally aligned observational geometry) 0.11~nHz. This leads to $\beta$ values from 1 (when the geometry is favorable) to 0.25 (a reduction of 50\% once the square root is taken for Equation~\ref{eq:minimum_detectable_flux_density}) for the highest frequency in the survey.

Plugging these values into Equation~\ref{eq:minimum_detectable_flux_density}, we find that a null result over the entire campaign would rule out \ac{ASP} transmitters with minimum detectable flux densities of approximately 282--1709 in the ``best case'' and ``worst case'' scenarios. The best case occurs when a signal is located at the center of the primary beam and has a drift rate close to the threshold that narrowly avoids fine channel attenuation, while the worst case occurs when the signal is located at the off-beam (with primary beam attenuation) and has the highest drift rate in the survey. For context, a transmission de-drifted to the barycenter would have a minimum detectable flux density of 578~Jy even if located at the off-beam. The most powerful transmitter on Earth, the now-defunct planetary radar from the $\sim$300~m Arecibo telescope, had an \ac{EIRP} of $\sim$20~TW; we can thus place limits on Arecibo-scale transmitters with $\Delta\nu_t = 1$~Hz out to 32--79~lyr in this survey. 

ATA-ASP will set the first limits of any kind on the \ac{ASP} and the most robust continuous transmitter upper limit on the \ac{ETZ} (covering ~10\% of its total sky area), which complements the more sensitive, higher-frequency limits on 20 \ac{ETZ} targets from \citet{sheikh2020breakthrough}.

We can also use the methodology from \citet{wright2018much} to determine the fraction of the ``Cosmic Haystack'' that we will cover in the full ATA-ASP survey, given the parameters described in the previous section. We use provided Python calculation scripts from \citet{wright2018much} to calculate the haystack fraction using the default haystack bounds and search bounds as follows: 

\begin{itemize}
    \item A time per scan of 300 seconds
    \item A minimum frequency of 2.3~GHz
    \item A maximum frequency of 3.7~GHz
    \item A minimum detectable flux density of 282--1709~Jy for the fiducial transmitter signal width of $\Delta\nu_t$ = 1 Hz (578~Jy at worst if de-drifted to the barycenter)
    \item A channel bandwidth of 1.9~Hz
    \item A polarization fraction of 1 (indicating instrument sensitivity to both polarizations)
    \item A sky coverage of 6.8 square degrees --- the sky coverage value is obtained by multiplying the width of the highest frequency (i.e., smallest) synthesized beamsize by the 360$^{\circ}$ circumference of the ecliptic. Then, another factor of two is multiplied in to represent the second, also searched, off-beam.
\end{itemize}

From these input parameters, we find a total haystack volume of $5.1 \times 10^{-21}$---$3.1 \times 10^{-20}$. This volume is about equivalent to the haystack fraction calculated in the previous largest \ac{SETI} survey ever done with the \ac{ATA} by \citet{harp2016seti}, and is the same order-of-magnitude as other large modern search campaigns in similar frequency regions \citep[e.g., ][]{gray2017vla}. Recall as well that even a single additional beam or tuning will increase the sky coverage by 50\%, providing ample opportunity for growth of the haystack fraction over the duration of the survey.

\section{Pilot Observations}
\label{sec:pilot_obs}

To prepare for the full-scale \ac{ASP}, we designed a pathfinder survey of a few nights of data, which we ran in Summer 2023, with the 20 dish beamformer. In the end, certain decisions which were made for this pilot project (e.g., the frequency tunings) were changed for the final survey based on the results of the pilot. Therefore, we describe the \ac{ASP} setup as it was performed in Summer 2023 in Section~\ref{ssec:pilot_observing_parameters_and_data_analysis}, uniquely from the overall survey design in Section~\ref{sec:survey_design}.

\subsection{Observing Parameters and Data Analysis}
\label{ssec:pilot_observing_parameters_and_data_analysis}

For the Summer 2024 pilot program we set the central frequencies of the two frequency tunings to 1.4~GHz and 2.042~GHz, which provided continuous coverage from 1.064--2.378~GHz. We selected this range to coincide with the spin-flip frequency of hydrogen at 1420~MHz, which has been proposed as a magic frequency in the past \citep[in fact, as one of the earliest \ac{SETI} Schelling points][]{Cocconi&Morrison}. The chosen integration time and frequency resolution for this survey were 15~seconds and 1~Hz respectively. The data were taken for an observation time of 5 minutes per scan, with scans looped continuously over period of time the \ac{ASP} is visible at the \ac{ATA} (around 10~pm--4~am local time throughout the summer at \ac{HCRO}). We opted to use two beams as shown in Figure~\ref{fig:ASP_geometry}, with an on-beam pointed directly at the \ac{ASP} and an off-beam placed 5 on-beam widths away from our on-beam as a control. 

We collected $\sim$17 hours of observation of the ASP over the course of 3 nights, which constituted 32~TB of filterbank data. Each file is the same HDF5 ``data cube'' described in Section~\ref{ssec:observing_strats}, where each tuning is divided into seven distinct frequency datafiles. To identify potential narrowband, drifting, sky-localized \ac{ASP} beacons in these files, we used the \texttt{turboSETI} package \citep{enriquez2019turboseti} and \texttt{NBeamAnalysis} code (as described in more depth in Section~\ref{ssec:analysis_strats}). In this pilot, we selected a \ac{SNR} threshold of 10 and a maximum drift rate of 10 Hz/s for \texttt{turboSETI}.

General parameters for both the full ATA-ASP survey and the pilot observations in SEction~\ref{sec:pilot_obs} are summarized in Table~\ref{tab:survey_params}:

\begin{table}[ht!]
    \centering
    \begin{tabular}{c|c|c}
    \hline
    \textbf{Parameter}           &   \textbf{Value in Full ATA-ASP Survey}    &   \textbf{Value in Pilot}   \\
    \hline
    \rowcolor{gray!50}
    \multicolumn{3}{c}{\textbf{\textit{Survey}}}  \\
    Number of Dishes    &   28--42                          &   20                      \\
    Number of Nights    &   365                             &   3                       \\
    \rowcolor{gray!50}
    \multicolumn{3}{c}{\textbf{\textit{Data Product}}}  \\
    Frequencies         &   2356--3700~MHz                  &   1064--2378~MHz          \\
    Total Bandwidth     &   1400~MHz                        &   1400~MHz                \\
    Fine Channel Width  &   1.9~Hz                          &   1~Hz                    \\
    Coarse Channel Width&   0.5~MHz                         &   0.5~MHz                 \\
    Time Per Scan       &   300~s                           &   300~s                   \\
    Time Integrations   &   16.8~s                          &   15~s                    \\
    \rowcolor{gray!50}
    \multicolumn{3}{c}{\textbf{\textit{Signal Search}}}  \\
    Search Software     &   \texttt{bliss}                  &   \texttt{turboSETI}      \\
    \ac{SNR} Thresholds &   15                              &   10                      \\
    Drift Rate Range    &   $\pm$4~Hz/s                     &   $\pm$10~Hz/s            \\
    Signal Bandwidths   &   $\sim$2--19~Hz                  &   $\sim$1--9~Hz           \\
    \ac{SEFD}           &   272--291~Jy                     &   332--336~Jy             \\
    Minimum Detectable Flux Density & 282--1709~Jy  &  115--936~Jy  \\
    Minimum Detectable Flux at barycenter drift rate (W/m$^2$) & 3.9$\times10^{-23}$ & 1.2$\times 10^{-23}$ \\
    Maximum Distance to detect a $\Delta\nu_t = 1$~Hz Arecibo-equivalent & 32--79~lyr &  44--124~lyr  \\    
    \hline
    \end{tabular}
    \caption{General search parameters for the planned ATA-ASP campaign and the pilot survey. The ``Minimum Detectable Flux at barycenter drift rate (W/m$^2$)'' column is calculated for a signal with a 0.11 nHz normalized drift rate (the maximum from Earth's contribution for an otherwise drift corrected signal) corrected to the highest frequency in each column, resulting in a signal which smears though a number of channels equivalent to (i.e., has an effective width of) $\Delta\nu_t$=6.8~Hz for the full campaign and $\Delta\nu_t$=4~Hz for the pilot survey. As the detected drift rate will often be less than the maximum in the barycentered case, we instead calculate the Arecibo-equivalent distance with $\Delta\nu_t = 1$~Hz throughout this work.}
    \label{tab:survey_params}
\end{table}

We then generated plots of the 500 hits with the lowest \texttt{NBeamAnalysis} DOT scores on each of the three nights, visually inspecting them for signals-of-interest. Once \texttt{NBeamAnalysis} had the beam ratio option available in addition to the DOT score (later in its development), we ran the data through the pipeline again to calculate beam ratio. We then plotted hits from all three nights with beam ratio above a chosen threshold of 4 (as the pilot program used only 20 dishes, not 28) and visually inspected these 614 hits in addition to the original 1500. 

\subsection{Results and Learnings} 
\label{ssec:pilot_results}

\begin{figure}[ht]
    \centering
    \includegraphics[width=\linewidth]{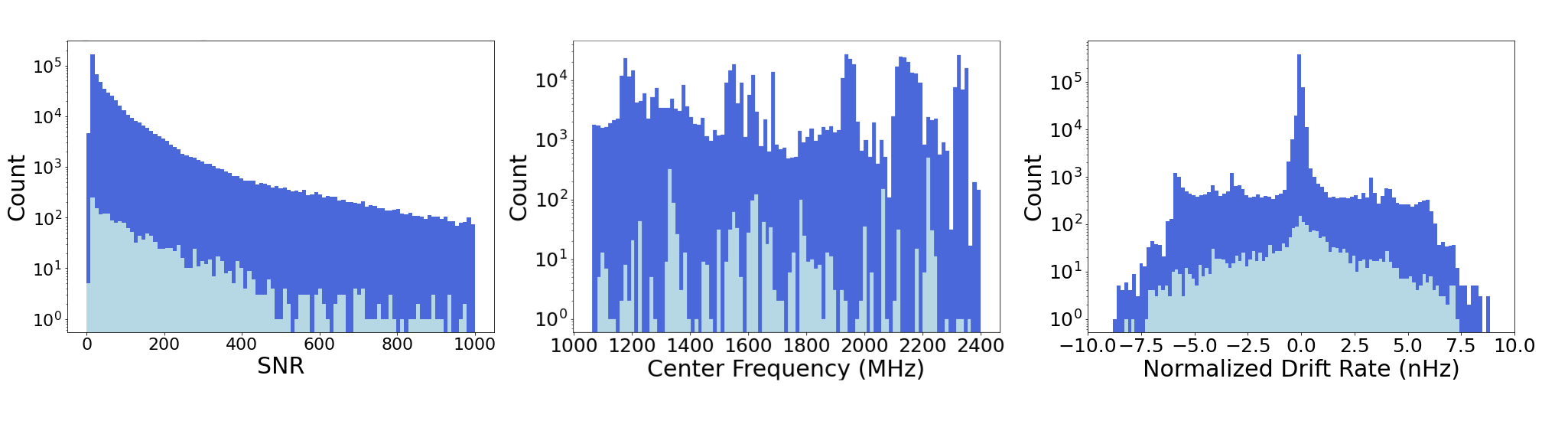}
    \caption{The characteristics of all hits in the \ac{ASP} pilot survey (dark blue) and the events that were visually inspected (light blue). Generally, the distributions of visually-inspected events trace the overall distribution in each variable. \textbf{Left: } Histogram of \ac{SNR} over the course of our three nights of observation. The lowest \ac{SNR} values for visually-inspected events on each day range from 10.00--10.25, indicating that the spatial filtering maintains the sensitivity from the initial minimum \ac{SNR} threshold of 10. \textbf{Center: } Histogram of center frequencies over the course of our three nights of observation. Certain frequency ranges, e.g., 1900--2000~MHz, are much more interference-heavy than others, as expected from frequency allocations in L-band. In addition, certain interference-heavy frequency ranges are more accurately removed by the \texttt{NBeamAnalysis} spatial filter, leading to the visually-inspected events not perfectly tracing the overall structure of the hits distribution.
    \textbf{Right: } Log-scaled histogram of normalized drift rates in nHz for all hits over the course of our three nights of observation. As expected, the majority of the hits are centered at zero drift, indicating terrestrial \ac{RFI} sources.}
    \label{fig:hit_characteristics}
\end{figure}

\texttt{turboSETI} identified 1.9$\times10^6$ narrowband hits within the 17 hours of data. \texttt{NBeamAnalysis} then removed 1.3$\times10^6$ as exact matches between the two beams, leaving us with 5.8$\times10^5$ hits which were initially ranked only on their DOT scores, as this was the only metric implemented in \texttt{NBeamAnalysis} at the time. The distributions for normalized drift rate, center frequency, and \ac{SNR} of both all hits and the visually-inspected events for all three nights are shown in Figure~\ref{fig:hit_characteristics}.

\begin{figure}[ht]
    \centering
    \includegraphics[width=0.8\linewidth]{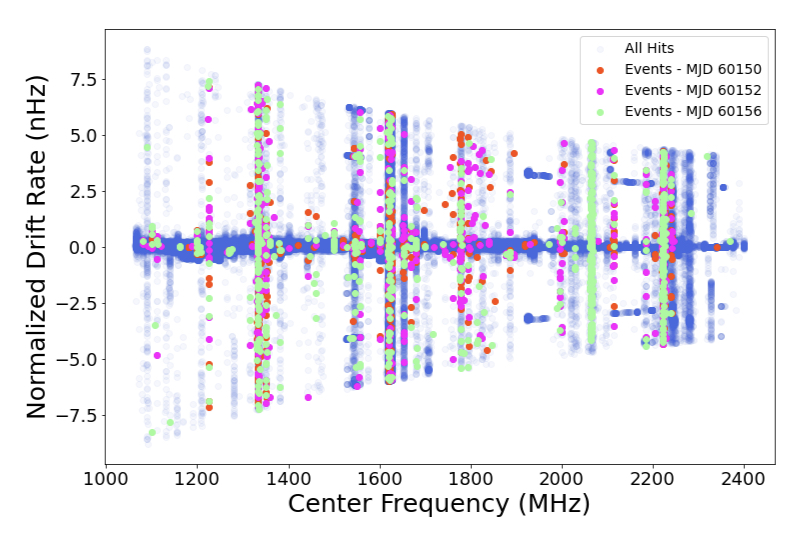}
    \caption{Center frequency plotted against normalized drift rate for all of our hits, as well as the plotted and visually-inspected events. Note that: a) most hits fall along a horizontal center line (i.e., their drift rates are near zero) b) some interference frequency ranges produce hits across all drift rates, not just zero c) the population tapers in drift rate space over center frequency --- this is due to the frequency-dependence of input, non-normalized drift rate versus the normalized plotting. Many plotted events with the lowest DOT scores still fall near zero drift rate, but some trace high drift interference ranges or, in some cases, appear in relatively empty parts of parameter space.}
    \label{fig:2d}
\end{figure}

Of the remaining 5.8$\times10^5$ hits, we generated dynamic spectra for visual inspection for the 500 hits with the lowest DOT scores on each night. We searched the 1500 produced plots for narrowband and drifting signals that were significantly stronger or exclusively present in the on-beam. The frequency and drift rate characteristics of the visually examined hits are shown in contrast to the entire hit population in Figure~\ref{fig:2d}. 

Figure~\ref{fig:2d} reveals some interesting patterns in how the hits with the lowest DOT scores fall in frequency and drift rate. Hits with the lowest DOT scores generally do not cluster around zero drift rate, unlike the overall hit distribution; this is sensible, given that hits at zero drift rate are likely to be ground-based \ac{RFI} entering through a sidelobe. \ac{RFI} of this type will be equally bright in the on- and off-beams, thereby warranting a high DOT score. Vertical lines in Figure~\ref{fig:2d} indicate frequency ranges with \ac{RFI} that triggers a \texttt{turboSETI} hit at every trial drift rate (usually broadband or morphologically complex \ac{RFI}). These lines are an artifact of the hit-finding process and also appear to a lesser degree in \texttt{bliss}. In this reduced view of the data, the most interesting hits are therefore the points in Figure~\ref{fig:2d} that do not fall along vertical or horizontal features but instead appear relatively isolated in the parameter space. Even so, the full population of highlighted hits in Figure~\ref{fig:2d} were visually inspected (not just the isolated points).

\begin{figure}[ht]
    \centering
    \includegraphics[width=1.0\linewidth]{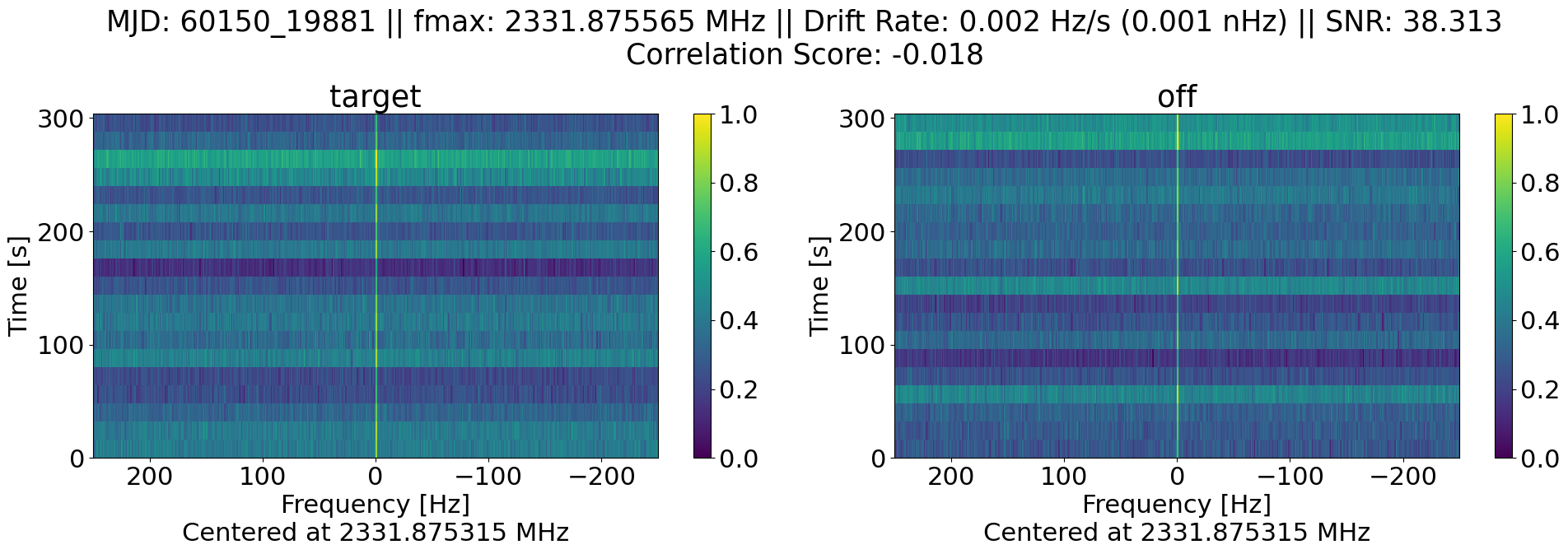}
    \caption{Event that was recorded by \texttt{turboSETI} and ranked as potentially sky-localized by \texttt{NBeamAnalysis} but was determined to be \ac{RFI} due to its nearly equal visibility in both beams and minuscule drift rate. This plot was representative of the majority of hits from our observations. It was likely flagged by \texttt{NBeamAnalysis} due to the changing broadband background noise level throughout the observation.}
    \label{fig:60150_19881}
\end{figure}

After visually inspecting our plots, we found no signals that were indicative of a radio technosignature from an \ac{ETA}. All of the hits produced by our analysis can be explained by terrestrial \ac{RFI}. The vast majority of signals plotted were narrowband, as expected, but were equally bright in both beams and displayed minimal drift rates, indicative of terrestrial \ac{RFI}. Figure \ref{fig:60150_19881} shows an example output dynamic spectrum from one of these signals.

\begin{figure}[ht]
    \centering
    \includegraphics[width=1.0\linewidth]{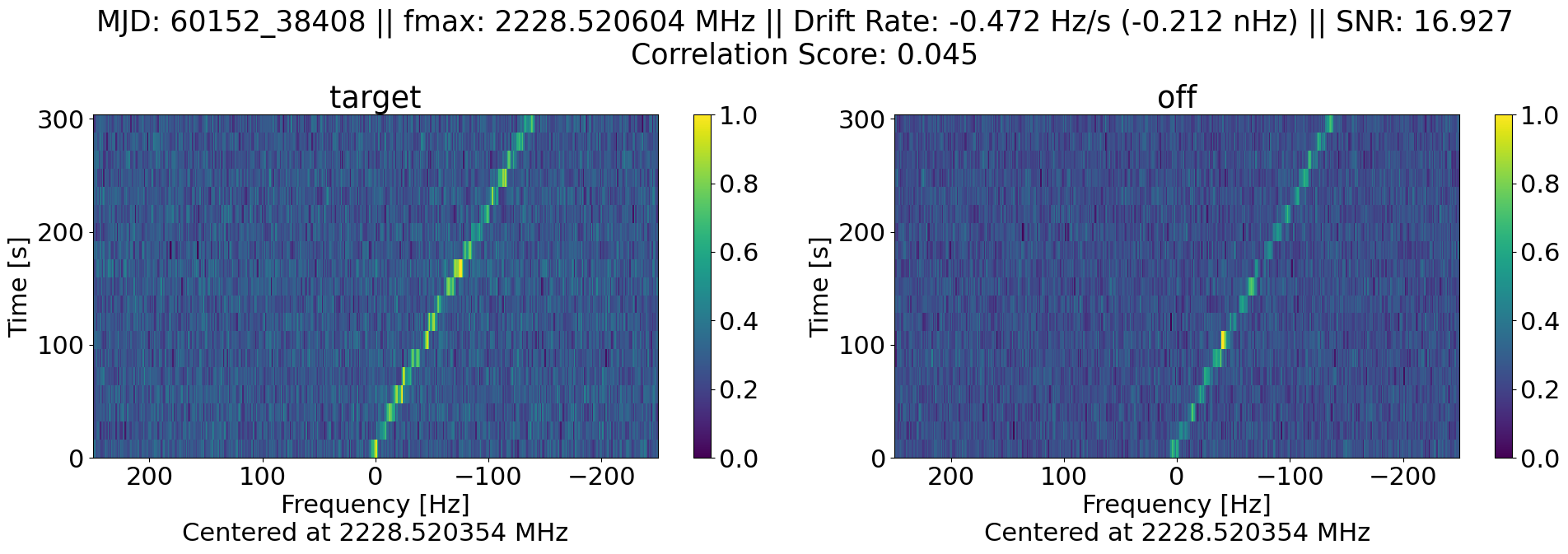}
    \caption{Drifting narrowband signal flagged by \texttt{NBeamAnalysis}. While the signal is definitively technological (narrowband) and drifting, its low beam ratio between the two beams and allocation are more consistent with \ac{RFI} from satellite communications or tracking.}
    \label{fig:60152_38408}
\end{figure}

Of the minority of signals that were both drifting and narrowband, none of them showed the expected characteristics of true sky-localization. Many of them were sets of aliases which appeared multiple times across the band, and were easily cross-referenced with known \ac{RFI} via the \ac{FCC} Table of Frequency Allocations. Figure \ref{fig:60152_38408} shows one of the ``most interesting'' signals in our sample, as it is an example of a narrowband, drifting, signal which is has somewhat different in \ac{SNR} between the on- and off-beams. After further consideration, this signal does not show a large enough beam ratio to be from a distant point source, and is most likely \ac{RFI} within the allocation used for ground-to-space, space-to-ground, and space-to-space transmissions for tracking, telemetry, and communications with satellites. 

\begin{figure}[ht]
    \centering
    \includegraphics[width=1.0\linewidth]{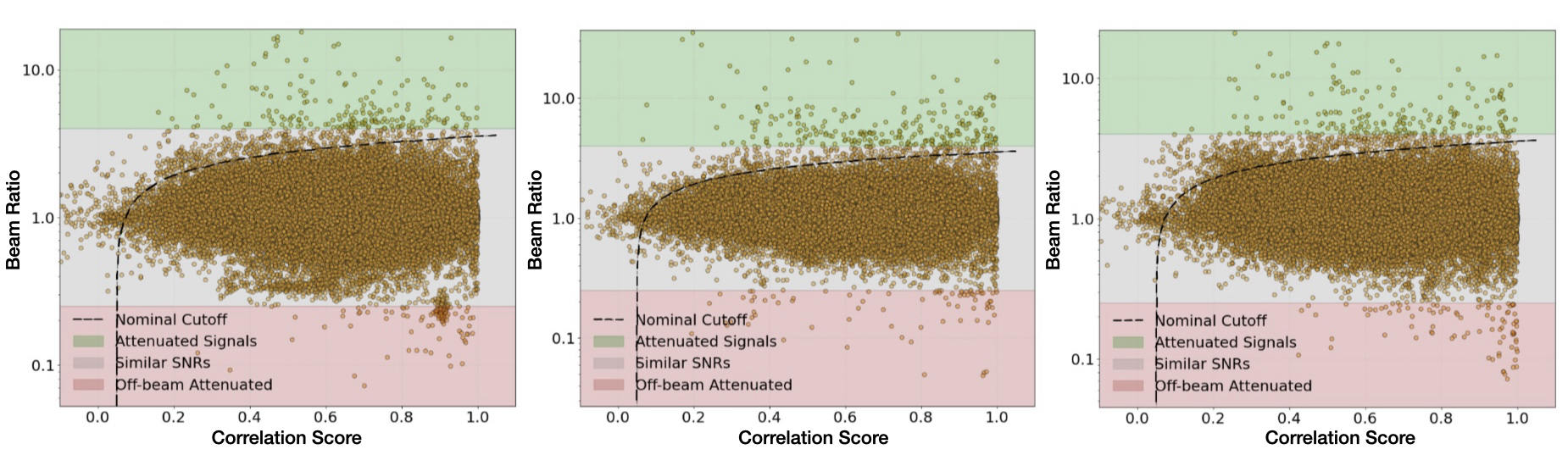}
    \caption{Population level plots for the three nights of data. On $x-axis$: DOT score. On $y-axis$: beam ratio. The grey and red regions represent hits with beam ratios below the chosen cutoff of 4. The nominal cutoff line is the cutoff for the 500 hits with the lowest DOT score, and the green region contains the hits with beam ratio above the cutoff of 4.}
    \label{fig:population-level}
\end{figure}

Figure \ref{fig:population-level} shows population-level plots for each of the three nights, with the dots in the green portion of the plot representing the hits above the beam ratio cutoff of 4 that were subsequently plotted.

Once again, all 614 plots generated in this second round of data analysis fell into the categories of noise or signals characteristic of ground-based RFI, leaving no signals of further interest.

We found no signals-of-interest in this pilot search of the \ac{ASP}, a null result which we can convert into an upper limit. The average \ac{SEFD} of the \ac{ATA} during our nights of observation was 272~Jy for our 1.4~GHz tuning, and 291~Jy for our 2.042 GHz tuning---we take an average value of 281.5~Jy for the following calculation. The minimum detectable flux density of a narrowband radio technosignature search is given in Equation~\ref{eq:minimum_detectable_flux_density}. The $\mathrm{SNR}_{threshold}$ for the pilot survey is 10, the channel width $\Delta \nu$ is 1~Hz, the number of polarizations $n_{pol}$ that the search is sensitive to is 2, and the observing time per scan is 5 minutes, with an assumed transmitter signal width of $\Delta f_t$ = 1~Hz. This formulation implies minimum detectable flux densities with $\Delta\nu_t=1$~Hz of 115--936~Jy for the same best and worst case scenarios, respectively, as discussed in Section~\ref{ssec:upper_limits}. This implies that an Arecibo-like transmitter with $\Delta\nu_t = 1$~Hz could be located as far away as 44--124~lyr and still be detectable in this survey. 

In conclusion, we took $\sim$17 hours of data, which represents 0.8\% of the full ATA-ASP survey described in Section~\ref{sec:survey_design}. From those hours of observation, we found 1.9$\times10^6$ total hits, and were left with 5.8$\times10^5$ hits after removing identical matches with \texttt{NBeamAnalysis}. We visually inspected the top 500 hits from each night and found 0 sky-localized signals-of-interest. From this, we can conclude that we find no narrowband technosignatures between RA range 20.266--21.312 on the \ac{ASP} above a flux density threshold of 115--936~Jy at these frequencies.

More generally, the pilot survey produced a series of ``lessons learned'' that influenced the full ATA-ASP survey design in Section~\ref{sec:survey_design}. 

\begin{enumerate}
    \item \textbf{Frequency Range: } The L-band frequencies in the pilot survey were extremely \ac{RFI}-dense, which caused the search algorithms to take an order-of-magnitude longer to run than would be needed for ATA-ASP. For the full survey, we opted for slightly higher frequencies (at the cost of a reduced coverage area) to avoid this issue.
    \item \textbf{\ac{RFI} Masking: } The pilot survey showed that we could identify certain ``problem'' allocations within the spectrum that were relatively narrow in bandwidth but would produce the majority of the hits in the survey. This inspired the \ac{RFI}-masking step that is being implemented on the full survey.
    \item \textbf{Beam ratio:} We found that the DOT score in \texttt{NBeamAnalysis}, while useful as a standalone metric, is more diagnostic of a signal's non-terrestrial origin if combined with a calculated beam ratio, as discovered in tests with Mars orbiters described in \citet{tusay2024ppo}. We are implementing the combined beam ratio plus DOT score filtering method in the full ATA-ASP survey.
    \item \textbf{Channel Width and Filesize: } Based on the datarate calculations from the pilot, we identified that slightly increasing the channel width and integration size would halve the filesize, allowing for more long-term storage options.
\end{enumerate}

\acresetall

\section{Conclusions and Future Work} 
\label{sec:conclusions}

The \ac{ASP}, a Schelling Point in time along the \ac{ETZ} bracketing the ecliptic, may be a particularly well-motivated space-time coordinate in the search for radio technosignatures. While the idea of searching for beacons in the \ac{ASP} has been present in the literature for some time, the actual observational search coverage is essentially zero. In this paper, we described a new campaign with the \ac{ATA}, titled ATA-ASP, which endeavors to perform the first large-scale search of this region in a multi-year, S-band observing campaign. We also presented a null result from pilot observations taken over three nights in Summer 2023, which demonstrated the technology and procedure for performing ATA-ASP and set the first upper limits on Arecibo-like transmitters with $\Delta\nu_t = 1$~Hz out to $\sim$44--124~lyr for the \ac{ASP}. When the full survey is completed, assuming two synthesized beams, we will have covered a haystack fraction of $5.1 \times 10^{-21}$---$3.1 \times 10^{-20}$ and placed limits on Arecibo-scale transmitters with $\Delta\nu_t = 1$~Hz out to 32--79~lyr.

As shown in Figure~\ref{fig:ASP_geometry}, the two-beam ATA-ASP survey, even when completed, will only cover $\sim$10\% of the \ac{ASP} in ecliptic latitude, or 20\% if limited to the restricted \ac{ETZ} \citep{heller2016search}. As such, we will endeavor to increase the latitude coverage as additional \ac{ATA} capabilities come online during the course of the survey. One option for expanding the latitude coverage is to add additional beams along the \ac{ASP} bisector, as shown in Figure~\ref{fig:ASP_geometry}. Currently, the corresponding increase in data rate and processing time is infeasible at these frequency resolutions and bandwidths: the frequency resolution would have to be increased by 250X and the bandwidth decreased by 6X to produce the $\sim$20 beams that would be needed to span the \ac{ASP}. However, as more compute power is added to the backend machines and the \texttt{NBeamAnalysis} code is optimized over the next few years, adding additional beams may be possible. A second option for expanding the latitude coverage is to perform the narrowband search within the ``image plane,'' i.e., to use the in-development BLADE correlator backend on the \ac{ATA} to create a sky-image of the entire \ac{ASP}. This spectral line correlator is still undergoing testing, and there are no existing open-source algorithms for performing such a search. However, as prototyping of image-plane narrowband search code begins over the next few years, it could be ported to a future epoch of ATA-ASP.

Thus far, we have taken a total of $>$121 nights of non-pilot data. Analysis of the first year of \ac{ASP} data will be presented in an upcoming publication.

\begin{acronym}
    \acro{SETI}{Search for Extraterrestrial Intelligence}
    \acro{ETA}{Extraterrestrial Agent}
    \acro{ETZ}{Earth Transit Zone}
    \acro{ASP}{Anti-Solar Point}
    \acro{ATA}{Allen Telescope Array}
    \acro{RFI}{Radio Frequency Interference}
    \acro{SNR}{Signal-to-Noise Ratio}
    \acro{FRB}{Fast Radio Burst}
    \acro{BLADE}{Breakthrough Listen Accelerated DSP Engine}
    \acro{FWHM}{full width at half maximum}
    \acro{BL}{the Breakthrough Listen Initiative}
    \acro{HCRO}{Hat Creek Radio Observatory}
    \acro{GBT}{Green Bank Telescope}
    \acro{FCC}{Federal Communications Commission}
    \acro{SEFD}{System Effective Flux Density}
    \acro{EIRP}{Effective Isotropic Radiated Power}
    \acro{DFT}{discrete Fourier transform}
\end{acronym}

\begin{acknowledgements}
    The Allen Telescope Array (ATA) refurbishment program and its ongoing operations have received substantial support from Franklin Antonio. Additional contributions from Frank Levinson, Greg Papadopoulos, the Breakthrough Listen Initiative and other private donors have been instrumental in the renewal of the ATA. Breakthrough Listen is managed by the Breakthrough Initiatives, sponsored by the Breakthrough Prize Foundation. The Paul G. Allen Family Foundation provided major support for the design and construction of the ATA, alongside contributions from Nathan Myhrvold, Xilinx Corporation, Sun Microsystems, and other private donors. The ATA has also been supported by contributions from the US Naval Observatory and the US National Science Foundation. S.Z.S. acknowledges that this material is based upon work supported by the National Science Foundation MPS-Ascend Postdoctoral Research Fellowship under Grant No. 2138147. E.H. participation was enabled through the SETI Institute REU program funded under NSF award No. 2051007.
\end{acknowledgements}

\bibliography{bibliography}
\bibliographystyle{aasjournal}

\end{document}